\documentclass[a4paper,fleqn,usenatbib]{mnras}

\usepackage{newtxtext,newtxmath}

\usepackage[T1]{fontenc}
\usepackage{ae,aecompl}

\usepackage{graphicx}	
\usepackage{amsmath}	
\usepackage{amssymb}	

\usepackage{bbm}
\usepackage{multirow}

\newcommand{\RR}{{\mathbb R}}
\newcommand{\cL}{{\mathcal L}}
\newcommand{\vepsilon}{\varepsilon}

\title[MAP estimation through SA for orbit determination]{Maximum a posteriori estimation through simulated annealing for binary asteroid orbit determination}

\author[I.D. Kovalenko et al.]{
Irina D. Kovalenko,$^{1,4}$\thanks{E-mail: irina.kovalenko@obspm.fr}
Radu S. Stoica,$^{2,1}$
Nikolay V. Emelyanov$^{3,1}$
\\
$^{1}$IMCCE, Observatoire de Paris, PSL Research University, CNRS,  Sorbonne Universit\'es, UPMC Univ. Paris 06, Univ. Lille,\\ 77 av. Denfert-Rochereau, 75014, Paris, France\\
$^{2}$Universit\'e de Lorraine, Institut Elie Cartan de Lorraine, 54506 Vandoeuvre-l\`es-Nancy Cedex, France\\
$^{3}$M. V. Lomonosov Moscow State University - Sternberg Astronomical Institute, 13 Universitetskij prospect, 119992 Moscow, Russia\\
$^{4}$Space Research Institute, Russian Academy of Sciences, 13 Profsoyuznaya ul. 84/32, 117342 Moscow, Russia
}

\date{Accepted 2017 July 21. Received 2017 July 19; in original form 2016 July 24.}

\pubyear{2017}

\begin{document}
\label{firstpage}
\pagerange{\pageref{firstpage}--\pageref{lastpage}}
\maketitle

\begin{abstract}
This paper considers a new method for the binary asteroid orbit determination problem. The method is based on the Bayesian approach with a global optimisation algorithm. 
The orbital parameters to be determined are modelled through an \textit{a posteriori} distribution made of \textit{a priori} and likelihood terms. The first term constrains the parameters space and it allows the introduction of available knowledge about the orbit. The second term is based on given observations and it allows us to use and compare different observational error models. Once the \textit{a posteriori} model is built, the estimator of the orbital parameters is computed using a global optimisation procedure: the simulated annealing algorithm. The maximum \textit{a posteriori} (MAP) techniques are verified using simulated and real data. The obtained results validate the proposed method. The new approach guarantees independence of the initial parameters estimation and theoretical convergence towards the global optimisation solution. It is particularly useful in these situations, whenever a good initial orbit estimation is difficult to get, whenever observations are not well-sampled, and whenever the statistical behaviour of the observational errors cannot be stated Gaussian like.
\end{abstract}




\section{Introduction}
The estimation of orbital parameters from observations is a classical inverse problem of celestial mechanics and it has been developed extensively for single asteroids. For binary asteroid mutual orbits, the problem is made complicated by the larger number of unknown parameters, as the mass of the system is usually unknown, and also by the insufficient observational data.

Several methods have been developed for binary objects mutual orbit determination. The classical least-squares orbit fitting to observations has been used  for satellites of asteroids \citep{Descamps} as well as for satellites of planets \citep{Emelyanov2005}. For binary  transneptunian objects \cite{noll2004orbit} used the downhill simplex Nelder-Mead algorithm \citep{nelder1965simplex}. 
\cite{vachier2012determination} developed a statistical method for mutual orbit fitting using a genetic-based algorithm.
All these methods are extremely efficient from a numerical point of view for the orbit determination problem, and give accurate results, but they require good initial estimations of the involved model parameters. Otherwise, the algorithms converge towards local optima of the fitting parameters.

Bayesian \textit{a posteriori} probability density formulation into orbit determination was introduced by~\cite{Muinonen}. Their approach was implemented in the so-called statistical ranging method by \cite{virtanen}, which uses Monte Carlo techniques for asteroid heliocentric orbit determination. The statistical ranging method constrains a volume of orbits in the orbital-element phase space, using the sampling in the observational space. Namely, each sample orbit is produced from two randomly chosen observations, introducing random deviations in the observations. Finally, a large set of orbits compatible with the observations is produced, while the six dimensions element space is constrained. 
\cite{oszkiewicz} developed the orbit ranging method that uses a Markov-Chain Monte Carlo (MCMC) method based on the Metropolis-Hastings algorithm. Another MCMC method, called virtual-observation MCMC method, was proposed by \cite{Muinonen201215}. In this method, the sampling is done in the parameters space instead of the orbital one. Thus, each sample orbit is generated from virtual observations with the downhill simplex Nelder-Mead algorithm.  

For binary mutual orbit determination \cite{grundy200842355} used the aforementioned Monte-Carlo techniques \citep{Muinonen,virtanen} in addition to the downhill simplex algorithm \citep{nelder1965simplex}.
\cite{hestroffer2005orbit} and \cite{oszkiewicz2013} developed techniques for binary asteroids based on the Monte Carlo sampling with the Thiele-Innes method \citep{aitken}. The Thiele-Innes method, initially proposed for binary stars orbit determination, requires the orbital period and three observations in the same tangent plane. \cite{oszkiewicz2013} use these parameters as the sampling parameters in the Metropolis-Hastings algorithm, so that the orbit sample is produced varying three observations in the same tangent plane, randomly selected from the whole set of observations, and orbital period. Even though the approach of \cite{oszkiewicz2013} has already been successfully applied to a few binaries, generally the use of the Thiele-Innes method may be inefficient for binary asteroids. The inherent parallax, due to the relative motion between the observer and the target, cannot be neglected, even for observations on close tangent planes, and the random selection of three observations in the same tangent plane usually can be not possible for a given dataset.

The method we propose in this paper continues and develops the previous ideas of Bayesian modelling. The \textit{a posteriori} model contains two terms: the likelihood and the prior. The likelihood term relies on the orbital parameters with the observations. In this work the likelihood is built so that different observational error distributions can be considered, whereas most methods usually assume only the Gaussian observational errors. The prior allows us to introduce an available knowledge about the orbit and to constrain the parameters space. By analogy with the optimisation problem it plays the role of a regularisation term. This approach is useful in case of newly discovered binary asteroids for preliminary orbit determination, when the initial parameters estimation is difficult.
Once the  likelihood and prior are built, the orbital parameters estimate is given by the {\it Maximum A Posteriori} (MAP). The MAP estimate is computed using the simulated annealing (SA) algorithm. This algorithm is a global optimisation method that guarantees convergence towards the global optimum of a considered function, independently of the initial state. The orbit sampling in the SA method is performed by the Metropolis-Hastings algorithm through the orbital parameters variation. The advantage of such an approach, over the sampling through observations, is  the avoiding parameters computation from observations (e.g. by the Thiele-Innes method) for each orbit sample.

The structure of the paper is as follows. In Section \ref{sec:dynammodel} the dynamical model with the binary asteroids observations are described. Then the problem in the form of a non-linear regression model is presented in Section \ref{sec:statmodel}. The least squares estimator and the new one we propose, associated with the simulated annealing algorithm, are described in Sections \ref{sec:leastsquares} and \ref{sec:simulatedanealing}, respectively. The new method validation on simulated and real observations is described in Section \ref{sec:practical}. Finally, conclusions and perspectives are given in Section \ref{sec:conclusions}.

\section{Dynamical model}
\label{sec:dynammodel}
\paragraph*{Binary system.}
A binary asteroid is a system of two asteroids orbiting their common centre of mass. Here, it is assumed that the motion of such systems follows the Kepler's laws, and the attractive force in the binary system is a gravitational force expressed by Newton's law. 
\paragraph*{Observations and coordinates.}
Unless the relative masses of two components are known, it is impossible to determine the centre of mass and therefore the orbits around the centre. But we are able to derive the Keplerian relative orbit of one component described about the other, which is assumed to remain stationary at the focus.

The motion that we observe is not the true motion but its projections on to planes perpendicular to the line of sight. The observation gives us the apparent position of one asteroid, called companion or secondary, with respect to another one, called primary, which is assumed to be at the focus. Every complete observation of a visual binary asteroid supplies with the time of observation $t$ and two coordinates on the sky plane. Let us set rectangular coordinates $x_t, y_t$ measured on the tangent plane. The position of the tangent plane is determined by $\alpha$ and $\delta$, right ascension and declination, respectively, at a given time $t$. Measured in arc seconds, the tangent coordinates are related to $\alpha, \delta$ through the equation: $x_t = (\alpha_2-\alpha_1)\cos\delta_1$ , $y_t = \delta_2-\delta_1$, where $(\alpha_1, \delta_1)$ and $(\alpha_2, \delta_2)$ are refereed to primary and secondary, respectively.

We introduce a coordinate frame, related to the tangent plane and centred at the primary asteroid. The $y$-axis is directed to the North, the $x$-axis to the Est and the $z$-axis is normal to the tangent plane and directed away from observer. In this frame the relative position of the secondary asteroid with respect to the primary one is related to observed positions through: $x = R x_t $ and  $y = R y_t$, where $R$ is the distance from the observer to the asteroid and the coordinates $x_t, y_t$ are transformed from arc seconds to radians.
 
Let us introduce an equatorial coordinate system $(x_E, y_E, z_E)$ referred to the epoch J2000 with the centre at the primary asteroid. Thus the rotation matrix from the tangent frame $(x, y, z)$ to the equatorial frame $(x_E, y_E, z_E)$ is
\begin{equation*}
S(\alpha,\delta) =
 \begin{pmatrix}
-\sin \alpha & -\cos \alpha \sin \delta & \cos \alpha \cos \delta \\
 \cos \alpha & -\sin \alpha \sin \delta & \sin \alpha \cos \delta \\
0 & \cos \delta & \sin \delta 
 \end{pmatrix}. 
 \end{equation*}
The topocentric distance $R$ as well as the
right ascension and declination $\alpha$, $\delta$ of an asteroid
at the moment of observation can be found
from ephemerides of asteroid heliocentric motion.
They can be treated here as known values.

\paragraph*{Time delay of the light.}
The time-light delay between observation and asteroid position has to be taken into account and  calculated by the linear equation $t = t_O-R/c$, where $t_O$ is a time of observation, $R$ is a distance between observer and target and $c$ is the light velocity.

\paragraph*{Orbital parameters.}
\label{sec:orbparams}
The keplerian orbit of the secondary asteroid with the primary being located in the focus of the ellipse is considered. This orbit is described by seven parameters $(a, e, i, \Omega, \omega, \tau, P)$: the six keplerian elements -- semi-major axis $a$, eccentricity $e$, inclination $i$, longitude of the ascending node $\Omega$, argument of pericenter $\omega$, time of pericenter passage $\tau$ -- and orbital period $P$, which is also an independent parameter under the condition that the mass of the system is also unknown. Otherwise, it can be derived using the third Kepler's low. The angles $ i$, $\Omega$, $\omega$ are referenced to J2000 equatorial frame.

\paragraph*{Coordinates from orbital parameters.}
\label{coordFromOrb}
The direct problem is to calculate the position of an asteroid at a given time from a known orbit. When the elements $(a, e, i, \Omega, \omega, \tau, P)$ are known, the apparent position coordinates $(x, y)$ at a given time $t$ are derived from the following equations:
\begin{equation*}
M = n(t-\tau), \quad  \hbox{where} \quad n = 2 \pi /P.
\end{equation*}
The eccentric anomaly $E$ is determined from the Kepler's equation: $M = E - e \sin E$.

The spherical coordinates, the radius vector $r$ and the true anomaly $\nu$, are calculated using the following equations:
\begin{equation*}
r = a(1-e \cos E); \qquad \tan \nu = \dfrac{\sqrt{1-e^2}\sin E}{\cos E - e}.
\end{equation*}
The coordinates in the equatorial frame are 
\begin{equation*}
\begin{array}{ll}
x_E = r(\cos u \cos \Omega-\sin u \sin \Omega\cos i), \\
y_E = r(\cos u \sin \Omega+\sin u \cos \Omega\cos i), \\
z_E = r \sin u \sin i,
\end{array}
\end{equation*}
where $u = \nu + \omega$ is argument of latitude.

The final $(x, y)$ apparent position coordinates are
\begin{equation*}
\begin{array}{ll}
x=- x_E\sin \alpha  + y_E\cos \alpha,\\
y=-x_E \cos \alpha\sin\delta- y_E\sin \alpha\sin\delta+z_E\cos\delta.
\end{array}
\end{equation*} 
Once the $(x, y)$ coordinates are calculated for a given set of orbital parameters, the observed minus calculated residuals (O-C) can be computed for each observation.

\section{Regression model}
\label{sec:statmodel}
The regression model construction needs to relate observations and theoretical positions to each other. Let each observation at a given time $t^{(k)}$ consists of two measured values $x_k^o$ $y_k^o$, $k=1,2 \dots N$, where $N$ is a number of observations. Let us denote
\begin{equation*}
\begin{array}{ll}
\varphi_{2k-1} = x_k^o,\\
\varphi_{2k} = y_k^o,\\
t_{2k-1}=t_{2k}=t^{(k)}.
\end{array}
\end{equation*}
Thus, the set of observations can be described by the following vector:
\begin{equation*}
\begin{array}{ll}
\boldsymbol\varphi&=( x_1^o, y_1^o, x_2^o, y_2^o \dots, x_N^o, y_N^o )^T\\ 
&= ( \varphi_1, \varphi_2, \varphi_3 \dots, \varphi_{2N})^T.$$
\end{array}
\end{equation*}
With a given theoretical model and time $t$ of observation the measured values $x$, $y$ can be computed. Namely, the theory (see Section \ref{coordFromOrb}) provides two functions:
\begin{equation*}
x=\psi_x(t, \boldsymbol\theta), \qquad
y=\psi_y(t, \boldsymbol\theta),
\end{equation*}
with $\boldsymbol\theta = \{\theta_j\} $, $j=1\dots m$  model parameters.

For a set of moments of time $\boldsymbol t=\{t_i\}$ $(i=1, 2, \dots, 2N)$, where $t_i = t^{(k)}$, $k=i/2$ when $i$ is an even, and $k=(i+1)/2$ when $i$ is an odd, we denote
\begin{equation*}
\boldsymbol \psi = (\psi_1, \psi_2, \psi_3 \dots, \psi_{2N})^T,
\end{equation*}
where
\begin{equation*}
\psi_i = \left\{
                \begin{array}{ll}
                \psi_x(t_i, \boldsymbol \theta),  \hbox{when \textit{i} is an odd};\\
                \psi_y(t_i, \boldsymbol \theta),  \hbox{when \textit{i} is an even}.
                \end{array}
        \right.
\end{equation*}
The so-called observational equation, which relates observations and theoretical positions to each other, can be expressed as follows:
\begin{equation}
\boldsymbol\varphi = \boldsymbol\psi (\boldsymbol\theta) + \boldsymbol\varepsilon,
\label{eq:obsequation}
\end{equation}
where $\boldsymbol\theta $ is a vector of seven unknown parameters that describes the orbit (see Section~\ref{sec:orbparams}), and $\boldsymbol\varepsilon$ describes observational and theoretical errors, which are also unknown $\boldsymbol\varepsilon = (\varepsilon_{x 1}, \varepsilon_{y 1} , \dots,\varepsilon_{x N}, \varepsilon_{y N})^T$.
The equation \eqref{eq:obsequation} represents a non-linear regression model. Statistical inference about unknown parameters $\boldsymbol\theta$ and $\boldsymbol\varepsilon$ can be drawn from $\boldsymbol\varphi$. Two methods for the problem of finding $\boldsymbol\theta$ are described here : the classical least squares estimator, and the new MAP estimate based on the SA algorithm.

\section{Least squares method}
\label{sec:leastsquares}
Orbital parameters fitting using the algorithm of least squares approach is a commonly used method in celestial mechanics. Here the application of this method is described for the binary asteroid orbital parameters determination. 

Let us use the same designations as in the observational equation~\eqref{eq:obsequation}. If observational and theoretical errors are neglected, a conventional observational equation can be written as follows:
\begin{equation*}
\boldsymbol\varphi = \boldsymbol\psi ( \boldsymbol\theta), \qquad \hbox{or} \qquad
\varphi_i = \psi_i ( \boldsymbol\theta) \,\, (i=1,2, \dots, 2N),
\end{equation*}
where $\boldsymbol\theta =(a, e, i, \Omega, \omega, \tau, P )= \{\theta_j\} $, $j=1...7$ is the vector of the true parameters values. Let  $\boldsymbol\theta^0 = \{\theta_j\} $, $j=1...7$ be the vector of initial parameters. Thus, the correction $\Delta \theta_j$ between each initial and true $j^{th}$ parameter is $\Delta \theta_j = \theta_j - \theta_j^0$,
and the conventional observational equation is
\begin{equation}
\label{adjustement}
\varphi_i = \psi_i (\boldsymbol\theta^0 + \Delta \boldsymbol\theta), \qquad (i = 1,2, \dots , 2N),
\end{equation}
where $\Delta\boldsymbol\theta=\{\Delta \theta_j\}$,  $j=1..7$. The least squares method assumes the preliminary orbital parameters estimation to be close enough to the final solution, so that it is used rather to adjust parameters than to determine them. Hence, it allows us to assume the $\Delta \theta$ to be small and to expand ~\eqref{adjustement} to the Taylor series. Let us retain only first-order terms of the $\Delta \theta_j$, thus
\begin{equation}
\label{conditeq}
\Delta \varphi_i = \sum \limits_{j=1}^7  J_{i,j} \Delta\theta_j \qquad (i = 1,2, \dots , 2N),
\end{equation}
where we denote $
\Delta \varphi_i = \varphi_i-\varphi_i^{c(0)}$, $\varphi_i^{c(0)} = \psi_i ( \boldsymbol\theta^0)$ and $J_{i,j} = \frac{\partial \psi_i(\boldsymbol\theta^0)}{\partial \theta_j}$.
The approximation~\eqref{conditeq} is the so-called conditional equation. These conditional equations are approximate, because, firstly, the errors of the theoretical model as well as the observational errors are neglected here, secondly, so do all terms higher than first-order terms in the Taylor series. Once the adjustments $\Delta\theta_j$ are derived, they are added to the initial parameters. This parameters determination method is called differential adjustment. The adjustments can be repeated many times, and, if the process converges, namely if adjustments become smaller, the algorithm can be stopped as soon as they are smaller than the assumed uncertainties.

At each step of parameters adjustment, the residuals of the conditional equations~\eqref{conditeq} are given by
\begin{equation*}
s_i = \Delta \varphi_i - \sum \limits_{j=1}^7  J_{i,j} \Delta\theta_j, \qquad (i = 1,2, \dots , 2N). 
\end{equation*}

There are many algorithms of parameters fitting. Here we use, the mean squares fitting that minimizes the sum of the squared residuals. To minimize the sum of squares of $s_i$, the gradient equation is set to zero and solved for $\Delta \theta_j$. A number of $7$ simultaneous linear equations is obtained. They are called the normal equations and written in the matrix notation as follows:
\begin{equation*}
\left(\boldsymbol{J^T}\boldsymbol{J}\right)\Delta \boldsymbol\theta=\boldsymbol{J^T}\Delta \boldsymbol\varphi.
\label{eq:norm}
\end{equation*}

\subsection{Weighted least squares}
The classical least squares method can be generalised by the weighted least squares. This method considers the case when observations have different variances by introducing weights.
Thus, if the measurements are uncorrelated, the normal equations should be modified as follows:
\begin{equation*}
\left( \boldsymbol{J^TWJ} \right)\Delta \boldsymbol\theta=\boldsymbol{J^TW}\Delta \boldsymbol\varphi,
\label{eq:normW}
\end{equation*}
where $\boldsymbol{W}$ is a diagonal weighted matrix, with $W_{i,i} = \dfrac{1}{\sigma_i^2} $, where $\sigma_i^2$ is a variance estimate of the $i^{th}$ measure.

\section{Bayesian modelling and simulated annealing}
\label{sec:simulatedanealing}

According to the Bayesian framework, the {\it a posteriori} probability density function of the unknown parameters $\boldsymbol\theta$ given the observed data  $\boldsymbol\varphi$ writes as
\begin{equation}
p(\boldsymbol\theta|\boldsymbol\varphi) \propto \cL(\boldsymbol\theta,\boldsymbol\varphi)\, p(\boldsymbol\theta),
\label{eq:model}
\end{equation} 
with $\cL(\boldsymbol\theta,\boldsymbol\varphi)$  and $p(\boldsymbol\theta)$, the likelihood and the {\it a priori} terms, respectively. The role of the likelihood term is to find the ``best'' orbital parameters fitting the observations. Nevertheless, using the likelihood term only, it will give an ill-conditioned problem. This drawback is solved by using a regularisation term, the prior density. 

Hence, under these considerations, the parameter estimator we propose is the {Maximum A Posteriori}, given by
\begin{equation}
\hat{\boldsymbol\theta} = \arg\max_{\theta \in \Theta}~p(\boldsymbol\theta|\boldsymbol\varphi),
\label{eq:map}
\end{equation}
where $\Theta \subset \RR^{7}$ is the parameters space.

\subsection{Likelihood model: conditional data term}
Hereafter, for the sake of simplicity and in agreement with the observational equation \eqref{eq:obsequation}, it is convenient to consider $N$ dimensional vectors instead of $2N$. Thus, $
\boldsymbol\varphi=\{\varphi_i\}, \, (i = 1,2, \dots , N)$, where $\varphi_i=( x^o_i, y^o_i)$ is the observed position of the secondary asteroid with respect to the primary, $t =(t_1, t_2, \dots, t_N)$, $t_i$ is a time of observation, $\boldsymbol\varepsilon = (\varepsilon_{x 1}, \varepsilon_{y 1}; \varepsilon_{x 2}, \varepsilon_{y 2}; \dots;\varepsilon_{x N}, \varepsilon_{y N})^T$ are the errors and $\boldsymbol \psi = \{x_i, y_i\}, \, (i = 1,2, \dots , N)$ is the set of theoretical positions. As before, the vector of unknown parameters is $\boldsymbol\theta =(a, e, i, \Omega, \omega, \tau, P )$.

For the construction of our model no particular assumption is favoured. Nevertheless, it is reasonable to assume that the observations are fitted as closely as possible by the optimal orbital parameters. One way to obtain this fitting is to allow the likelihood to depend on the distances from observed points to their associated calculated points. The calculated point is the point position computed using the current orbital parameters at the same time $t$ as the given observed point.

The likelihood function in~\eqref{eq:model} is defined by
\begin{equation*}
\cL(\boldsymbol\theta,\boldsymbol\varphi) = \prod_{i=1}^{N}p(\varphi_i|\btheta),
\end{equation*}
where we set
\begin{equation*}
p(\varphi_i|\btheta) =  \exp \left[ - \left( |x^{O}_i-x^C_i|^l +  |y^O_i-y^C_i|^l \right)^{k/l} \right] = \exp[ - \vepsilon_i(l)^{k}],
\end{equation*}
with $(x_i^{o},y_i^{o})$ and $(x_i^{c},y_i^{c})$ being the coordinates of observations and the computed positions at the same associated times, respectively. The functions $\vepsilon_i(l)$ may be interpreted as the equivalent of the residuals in a regression model. For instance, whenever $(k,l)=(1,1)$, the residuals tend to exhibit a Laplacian character, while for $(k,l) = (2,2)$, a Gaussian one.

Furthermore, whenever the uncertainties of observations are available, the likelihood writes as
\begin{equation}
\cL(\boldsymbol\theta,\boldsymbol\varphi) =  \exp\left[ - \sum_{i=1}^{N}  \vepsilon_{i}(l)^{k} \right],
\label{eq:likelihood}
\end{equation}
while the residuals are given by
\begin{equation*}
\vepsilon_{i}(p)=\left( (|x^{O}_i-x^C_i|/\sigma_{x,i})^l + (|y^O_i-y^C_i|/\sigma_{y,i})^l \right)^{k/l}, \quad (i = 1,\ldots, N)
\end{equation*}
with $\sigma_{x,i}$ and $\sigma_{y,i}$ the estimated uncertainties for a given observation.

\label{sec:likelihood}
In this work the following four likelihood models \eqref{eq:likelihood} are compared: 
\begin{itemize}
\item model 1: $(k, l) = (1, 1)$, $\sigma_{x,i}=\sigma_{y,i} = 1$, $(i=1, 2, ..., N)$;
\item model 2: $(k, l) = (2, 2)$, $\sigma_{x,i}=\sigma_{y,i} = 1$, $(i=1, 2, ..., N)$;
\item model 3: $(k, l) = (1, 1)$, $\sigma_{x,i}$, $\sigma_{y,i}$ $(i=1, 2, ..., N)$ are the given uncertainties, estimated for each observation;
\item model 4: $(k, l) = (2, 2)$, $\sigma_{x,i}$, $\sigma_{y,i}$ $(i=1, 2, ..., N)$ are the given uncertainties, estimated for each observation.   
\end{itemize}

\subsection{A priori model: regularisation term}
\label{sec:initialisation}
The \textit{a priori} model allows the introduction of available knowledge regarding the joint probability distribution of the orbital parameters. For instance, if correlations between the different parameters are previously known, they can be introduced through the prior model. This term also defines the restriction of the parameter space to those of the considered celestial bodies. 

Whenever no particular information is available, a prior should have minimal influence on the inference. 
A possible choice is to use the Jeffreys \citep{jeffreys1946invariant} non-informative priors. Jeffreys noted that for independent parameters the prior should be treated separately, thus in our problem it is convenient to use
\begin{equation*}
p(\theta) = \prod \limits_{i=1}^7 p(\theta_i),
\end{equation*} 
where $\theta_i$ is one of the seven parameters describing the orbit.

\label{sec:prior}
Following the Jeffreys' principle of non-informative prior choice when the parameter space is a bounded interval \citep{jeffreys1973scientific}, we set the independent prior distributions for each parameter to be uniform. These distributions were defined over the largest interval of possible values corresponding to each parameter.

\paragraph*{Semi-major axis $a$.} 
The lower limit on the semi-major axis can be derived from observations. Let  $\rho$  be the observed distance between two components in a binary system. As a projection on to the tangent plane, the maximal value $\rho$ is never greater than the sum of the semi-major axis $a$ and the distance $c$ from the focus to the centre of the orbit ellipse:
$\rho_{max} \leq a + c$, where $c = ae$.
In the case of a very elongate orbit the eccentricity is $e \approx 1$, thus $\rho_{max} \leq 2a$. It allows us to set the lower limit on the semi-major axis: $a_{min}= 0.5\rho_{max}$. While the lower limit can be derived, an approximation is needed for the upper limit. Namely, the \textit{a priori of the semi-major axis is set to be} 
\begin{equation*}
a \thicksim \mathcal{U} [a_{min}, a_{max}],
\end{equation*}
where $a_{max}$ is the maximum semi-major axis among known binaries in the same dynamical group (e.g. near-Earth, main-belt, trans-Neptunian object). 
\paragraph*{Eccentricity $e$, longitude of the ascending node $\Omega$ and argument of periapsis $\omega$.}
In the absence of an \textit{a priori} information about the eccentricity $e$, the longitude of the ascending node $\Omega$ and the argument of periapsis $\omega$, we assume that they follow uniform distributions in the interval of all possible values:
\begin{equation*}
e \thicksim \mathcal{U}[0, 1],  \quad
\Omega \thicksim \mathcal{U}[0, 2\pi],
 \quad \omega \thicksim \mathcal{U}[0, 2\pi].
\end{equation*} 

\paragraph*{Inclination $i$.}
In the case when the topocentric direction of the asteroid
is constant two mirror orbits are possible: direct and retrograde. The first one has an inclination between 0 and 90 degrees, the second one between 90 and 180 degrees. Thus, the problem can be split for finding two solutions that correspond to the following \textit{a priori}:
\begin{equation*}
\begin{array}{ll}
i \thicksim \mathcal{U}[0,\frac{ \pi}{2} ], \hbox{direct motion}, \qquad \hbox{or} \\
i \thicksim \mathcal{U}[\frac{ \pi}{2} , \pi], \hbox{retrograde motion}.
\end{array}
\end{equation*}
In the case of distant asteroids or trans-Neptunian binaries (TNBs) the topocentric direction can be almost constant. In general case when the orbit is observed
from different sides (with different topocentric directions)
the inclination value can be distinguished from observations.

\paragraph*{Orbital period $P$.}
The interval of possible values for the orbital period cannot be constrained analytically from observations and/or dynamical model. In order to set a prior for this parameter, approximations were done based on the properties of the asteroid dynamical group. We denote the \textit{a priori} distribution
\begin{equation*}
P \thicksim \mathcal{U} [P_{min}, P_{max}],
\end{equation*}
where $P_{min}$ and $P_{max}$ are the minimal and maximal values of the orbital period among known binaries in the same dynamical group. Additionally, the average asteroid, the $a_{min}$ and the third Kepler's law can be used in order to restrict the prior interval. 

\paragraph*{Time of periapsis passage.}
The time of periapsis passage $\tau$ can be bounded by the moment of the first observation $t_{min}$. The \textit{a priori} distribution for $\tau$ is uniform in the range equals to the period:
\begin{equation*}
\tau \thicksim \mathcal{U} [t_{min}, t_{min}+P_{max}].
\label{T0apriori}
\end{equation*}
where $P_{max}$ is the upper orbital period \textit{a priori} limit.

\subsection{Optimisation algorithm: the simulated annealing}
\label{sec:SA}
For computing the MAP estimate~\eqref{eq:map}, a simulated annealing (SA) algorithm has been built. This algorithm is a general global optimisation technique that works by iteratively simulating $p(\bvarphi|\btheta)^{1/T}$ while slowly cooling the temperature $T$~\citep{kirkpatrick1983optimization}. The principle of the method is the following. At high temperatures all the configurations are accepted, the simulated probability distribution is equivalent to the uniform distribution over the entire configuration space. While the temperature goes down, the algorithm chooses those states that tend to maximize the considered probability distribution. If the temperature goes down slowly enough, then the algorithm makes its choices while getting out from the local optima. At convergence, when temperatures $T \rightarrow 0$, the algorithm is frozen at the desired global maximum. In theory, the algorithm converges weakly to the uniform distribution over the sub-space of configurations that maximize the probability density of interest~\citep{GemaGema84,StoiGregMate05}.

Two ingredients are needed to set-up such an algorithm. The first is a sampling algorithm from the considered probability distribution, and the second is a cooling schedule for the temperature parameter. For the sampling algorithm, we chose to build a Metropolis-Hastings (MH) dynamics~\citep{HASTINGS01041970}. This algorithm converges theoretically after an infinite number of iterations, and the speed of its convergence depends on the used proposal densities. 
The initial set of parameters $\boldsymbol\theta^{(0)}$ may be chosen according to the \textit{a priori} distribution. Being in the state $\theta^{(i-1)}$ a new parameter value $\theta^{\prime}$ is proposed using a uniform proposal density for each parameter
\begin{align}
Q(\boldsymbol\theta' |\boldsymbol\theta^{(i-1)}) & =\prod \limits_{j=1}^7 \dfrac{\mathbbm{1}\{\theta_j' \in [\theta_j^{(i-1)}-\Delta \theta_j, \theta_j^{(i-1)}+\Delta \theta_j]\} }{2 \Delta \theta_j}.
\label{eq:proposal}
\end{align}
On the one hand, if $\Delta \theta_j$ is too high, the new proposed state risks to be rejected. On the other hand, if $\Delta \theta_j$ is too low, the algorithm does no travel fast enough through the configuration space. For our problem the values for $\Delta \theta_j$ were found after several trials and errors. Here, since the parameter space is a compact in $\RR^d$, these choices for the proposal densities guarantee that the algorithm converges with equal speed independent of initial state \citep{RobeSmit94,RobeTwee96}.

\cite{GemaGema84} and \cite{StoiGregMate05} proved the existence of logarithmic cooling schedules that guarantee the convergence of their proposed SA algorithms associated to their corresponding simulation dynamics. Establishing such a result for any probabilistic model and simulation dynamics is a non-trivial task. From a more practical perspective, a common choice for the cooling schedule is 
\begin{equation}
\label{eq:temperature}
T_i = T_{\max}~c^i,
\end{equation}
with $T_{\max}$ being the initial temperature and  $c$ a constant in the interval $[0.95, 1)$. As an option, it may be preferable to slow down this exponential cooling schedule, while reducing the correlation between samples. This can be done by reducing the temperature every $m$ iterations, where $m$ is the so-called de-correlation step.

\subsection{Summary of the algorithm}
\label{sec:algorithm}
The implemented SA algorithm consists of the following steps: 
\begin{enumerate}
\item Initialisation
\begin{itemize}
\item Choose an initial value for the orbital parameters $\btheta_{0} \in \Theta$. The parameters are randomly chosen according to the prior distributions.
\item Set an initial temperature $T_{\max}$.
\item Set a counter $i=1$.
\item Set stop conditions: a limiting number of successive temperatures $n_{max}$ when any new proposal orbit is not accepted (the system is called  "frozen" \cite{li2009markov}) and a limiting number of iterations $i_{max}$.
\item Set a de-correlation step $m$.
\end{itemize}
\item 
For each iteration $i$ do
\begin{itemize}
\item Generate a candidate set of orbital parameters from the proposal distribution $$\boldsymbol\theta'\thicksim Q(\boldsymbol\theta' | \boldsymbol\theta^{(i-1)}),$$ 
where $Q(\boldsymbol\theta' | \boldsymbol\theta^{(i-1)})$ is given by~\eqref{eq:proposal}. 
\item
Compute the acceptance probability:
\begin{equation*}
\label{accoef}
\alpha = \min \left\{1,\left[\dfrac{p(\boldsymbol{\theta'|\varphi})}{p(\boldsymbol{\theta^{(i-1)}|\varphi})}\right]^{1/T_{i-1}}\right\}.
\end{equation*} 

\item
The new orbit is accepted with probability $\alpha$. 

In  practice,  a random  number  $\alpha_r$  is  generated  in  the  range  $[0,1]$  and  is compared  with $\alpha$. 
If $\alpha_r<\alpha$, the new orbit is accepted $\boldsymbol\theta^{(i)} = \boldsymbol\theta'$, otherwise it is rejected and the last accepted orbit is set as the current one  $\boldsymbol\theta^{(i)} = \boldsymbol\theta^{(i-1)}$.
\item
If $(i\, \text{mod} \, m) = 0$, set a new temperature $T_i = h(i, T_{\max})$, where $h$ is the cooling schedule~\eqref{eq:temperature}.
\item Set $i = i + 1$.
\end{itemize}
\item The algorithm stops if no new proposed orbit is accepted during $n_{max}$ successive changes of the temperature or if it achieves $i=i_{max}$.
\end{enumerate}

\subsection{Uncertainty estimation}
\label{sec:uncertEstimation}
In order to obtain a statistical evaluation of the result, the algorithm \ref{sec:algorithm} was performed $100$ times using different initial sets of orbital parameters $\btheta_{0}$, that were chosen following the prior distribution. The initial temperature $T_{\max}$ was the same for all of these situations. Thus, we obtain $100$ orbits from which we retain the one, described by the set of orbital parameters $\hat{\btheta}$, that  produces the greatest \textit{a posteriori} value.

The parameters uncertainty estimation is given by the empirical variance and an inter-quantile interval. The variance for each parameter is estimated from the obtained $100$ orbit samples. Then the uncertainty is given by two estimated standard deviations $\pm 2\sigma_\theta$. The length of the chosen inter-quantile interval of level 95$\%$ for each parameter is computed using the difference of the corresponding empirical quantiles $q(0.975)$ and $q(0.025)$. In addition, the length of this interval can be compared with $4\sigma_{\theta}$ as a verification of the Central Limit Theorem.

The approach of uncertainty estimation through inter-quantile intervals is more robust to those values that tend to be either too small or too large. This is not the case for the variance estimate, since this estimate depends on the sum of all of the considered values. Under these circumstances, for the ephemeris uncertainty estimation we use only the inter-quantile interval. The uncertainty for a predicted position at a given time \textit{t} is estimated by a 95$\%$ inter-quantile interval as follows. Once the sample of orbits is obtained, we calculate the sample of positions at time \textit{t}. For the obtained sample the 95$\%$ inter-quantile interval is obtained using the empirical quantiles $q(0.025)$ and $q(0.975$ for each position coordinates. If the coordinates of the observed position at the corresponding time $t$ are not inside the inter-quantile interval, the hypothesis that the observed position may be a realisation of the estimated model is rejected, with a $p-$value lower than $5\%$.

\section{Practical Orbit Computation}
\label{sec:practical}

\subsection{Implementation}

The input data consist of a set of observations, the likelihood and the \textit{a priori} models, and the SA configurations. For each  observation at time $t_i$ $(i=1, 2, ..., N)$ the following data should be provided:
the coordinates $x_i, y_i$ of the relative position of the secondary with respect to the primary, the topocentric distance $R_i$, and the right ascension $\alpha_i$ and declination $\delta_i$ of the primary in the geoequatorial frame of the J2000 epoch.

The SA algorithm parameters (the proposal distributions and the temperature schedule) were set after several trials and errors. The proposal distribution was chosen to be the uniform distribution with support $\Delta \theta_j= 0.1\, \Delta \theta_{(prior)j}$, where $\Delta \theta_{(prior)j}$ defines the associate uniform prior. There is a lot of freedom to choose the proposal distribution, guaranteeing the theoretical convergence of the SA, while preserving the uniform ergodicity of the simulated Markov chain at the basis of the optimisation algorithm. Nevertheless, choosing the optimal proposal distributions in order to get the highest speed of convergence is an open mathematical problem. Therefore, the previously mentioned choice was preferred in agreement with the theoretical requirements, but also due to its simplicity. The cooling is performed according to the formula \eqref{eq:temperature}, with de-correlation step $m=50$, coefficient $c=0.999$ and $T_{max}= 10^7$. The algorithm stops if any proposal orbit is not accepted at $100$ successive temperatures. 

Experiments on simulated and real data sets were conducted in order to verify the proposed method. The considered data sets are related to TNBs and are presented in Appendix \ref{app:A}.

In the following, the four likelihood models described in Section \ref{sec:likelihood} are analysed. The prior distribution is described in \ref{sec:prior}. In the particular case of TNBs, the maximum semi-major axis for the prior is given $a_{max}= 102000$ km \citep{Johnston}. The orbital period according to \cite{Johnston} is ranged from $\approx 0.5$ to $6300$ days. Since this interval is large, we restrict the prior and choose $P \thicksim \mathcal{U} [0.5, 1000]$. This estimation is in accordance with the average mass of known TNBs $m \approx 4.5 \times 10^{18}$ kg and the calculated $a_{min}$ (according to the third Kepler's law).

\subsection{Simulated observations}

\begin{figure}
\centering
\includegraphics[width=\linewidth]{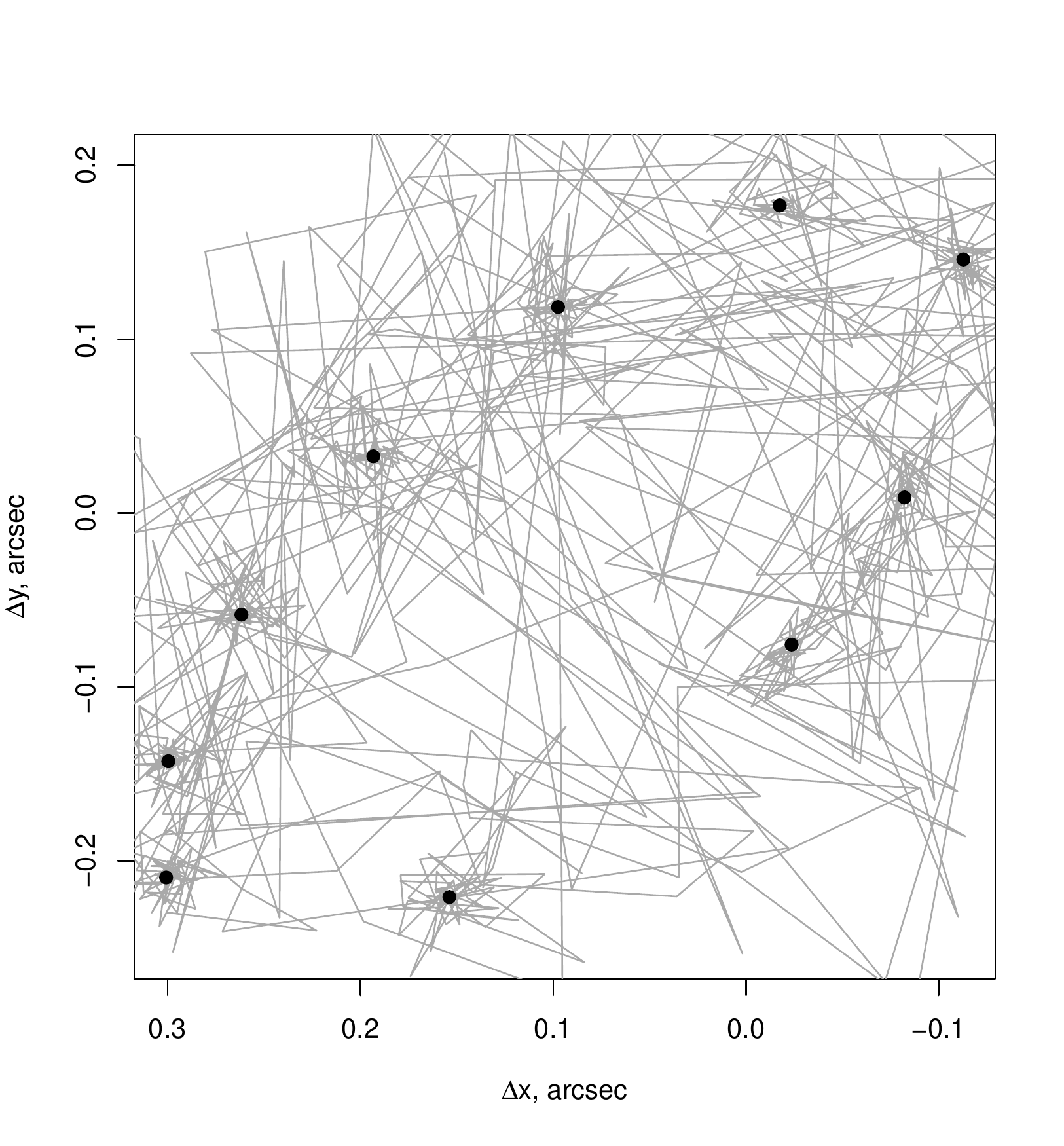}
\caption{Simulated observations (black points): $\Delta x$ and $\Delta y$ correspond to relative positions of the secondary with respect to the primary. Lines show a search for the optimal solution during SA algorithm (only each 100th change of calculated position, from the last 7000 accepted orbits, is shown).}
\label{fig:artobs}
\end{figure}

A number of $10$ artificial observations were generated using the model of the binary asteroid motion, described in Section $2$, and the following orbital parameters: $a = 10000$ km, $e=0.5$, $i = 135^o$, $\Omega = \omega = 45^o$, $\tau = 0$, $P = 30$ days. The right ascension $\alpha$, declination $\delta$ and geocentric distance correspond to those of the trans-Neptunian binary Altjira. Here we assume that the difference between the geocentric and topocentric $R$ distances is neglected. The values of coordinates $\alpha$, $\delta$ and $R$ are set from ephemerides of its heliocentric motion on the range between 2454000.5 JD (for the first observation) and 2454057.5 JD (for the last observation) (see Table \ref{tab:obssim}). The simulated observations cover one orbital revolution of the secondary around the primary. The choice of the distant trans-Neptunian object allows us to avoid a significant tangent plane changing during the considered interval of time and to graphically demonstrate the distribution of observations (see Figure \ref{fig:artobs}).        
The 10 generated observations of the relative positions, listed in Table \ref{tab:obssim}, are non-homogeneously distributed along the apparent orbit (see Figure \ref{fig:artobs}). We randomly set the observational uncertainties in the range from 0.001 to 0.01 arcsec following the uniform distribution (see Table \ref{tab:obssim}).

\paragraph*{Results.}
\begin{table}
\centering
\caption{Statistical summary of parameters distribution, obtained with the 100 SA algorithm evolutions applying to simulated observations. $\sigma_{\theta}$ denotes the empirical standard deviation. \textit{m.} denotes a likelihood model number. $\Delta q$ denotes $q(0.975)-q(0.025)$, namely 2.5$\%$ and 97.5$\%$ quantiles interval. a) Angles are referenced to J2000 equatorial frame. b) The mean anomaly is referred to the epoch 15 days.}
\medskip
\label{tab:resultSim}
\begin{tabular}{llllll}
Parameter     & m.  & mean $\pm 2\sigma_{\theta}$       & $\Delta q$        \\ \hline
$a$, km       & 1 & 10006.93 $\pm$ 197.86              & $710.27$                     \\
              & 2 & 9989.47 $\pm$ 97.75              & $341.65$                     \\
              & 3 & 9995.97 $\pm$ 132.85              & $516$                     \\
              & 4 & 9986.31 $\pm$ 93.55              & $380.68$                     \\ \hline
$e$           & 1 & $0.5001\pm0.02838$   & $0.05194$  \\
              & 2 & $0.4995\pm0.00604$   & $0.00992$  \\
              & 3 & $0.4987\pm0.02941$   & $0.05947$  \\
              & 4 & $0.4991\pm0.00281$   & $0.00562$  \\ \hline
$i^a$, deg      & 1 & $134.98\pm1.8$                       & $3.36$                     \\
              & 2 & $135.02\pm0.08$                       & $0.15$                     \\
              & 3 & $135.04\pm1.24$                       & $2.23$                     \\
              & 4 & $135.05\pm0.33$                       & $0.65$                     \\ \hline
$\Omega^a$, deg & 1 & $44.95\pm2.4$                     & $4.39$                   \\
              & 2 & $44.94\pm1$ & $2.1$                   \\
              & 3 & $44.94\pm2.87$                     & $5.51$                   \\
              & 4 & $44.92\pm1$ & $2.54$                   \\ \hline
$\omega^a$, deg & 1 & $45\pm2.01$                      & $3.64$                    \\
              & 2 & $45.04\pm0.4$  & $0.68$                    \\
              & 3 & $44.91\pm1.3$                      & $2.74$                    \\
              & 4 & $45.07\pm0.66$                      & $1.32$                    \\ \hline
$P$, days    & 1 & $30\pm0.52$                       & $0.93$                     \\
              & 2 & $29.99\pm0.43$                       & $0.75$                     \\
              & 3 & $29.99\pm0.44$                       & $0.93$                     \\
              & 4 & $29.99\pm0.4$  & $0.83$                     \\ \hline M$^b$, deg    & 1 & $180\pm7.93$                       & $14.42$                     \\
              & 2 & $179.86\pm5.19$                       & $9.04$                     \\
              & 3 & $180.25\pm5.12$                       & $9.6$                     \\
              & 4 & $179.78\pm5.34$                       & $10.92$                     \\ \hline
\end{tabular}
\end{table}

An example of the SA behaviour is shown on  Figure \ref{fig:artobs}. This plot shows how the calculated positions evolve towards the matching of the observations during the MAP search.

\begin{figure}
\centering
\includegraphics[width=\linewidth]{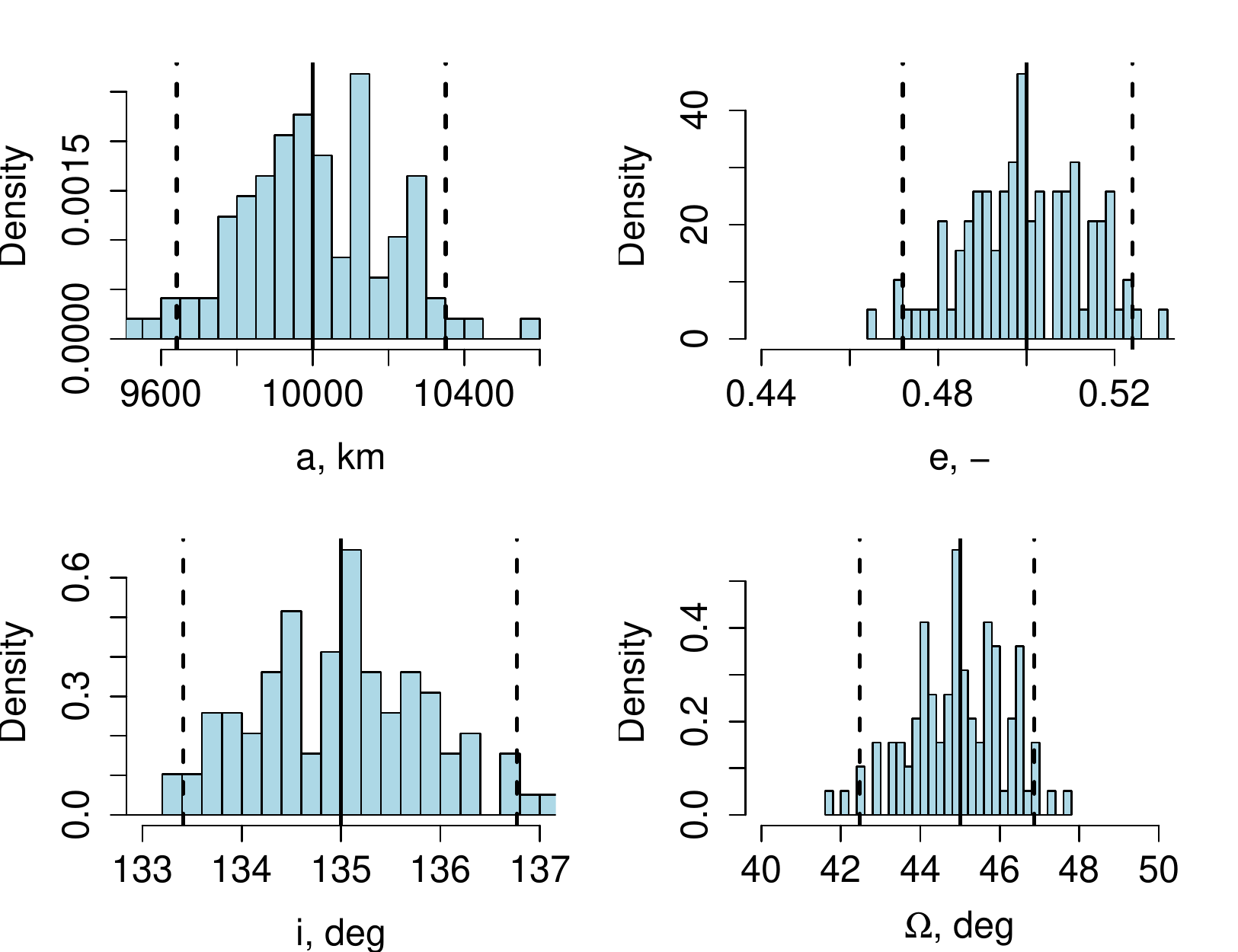}
\caption{Resulting distributions of semi-major axis $a$, eccentricity $e$, inclination $i$ and longitude of the ascending node $\Omega$ (referenced to J2000 equatorial frame) obtained for simulated observations with the likelihood model 1. Solid line represents the true parameter value. The doted lines represent the 2.5$\%$ and 97.5$\%$ quantiles of the  resulting sample.}
\label{fig:simMod11}
\end{figure}

\begin{figure}
\centering
\includegraphics[width=\linewidth]{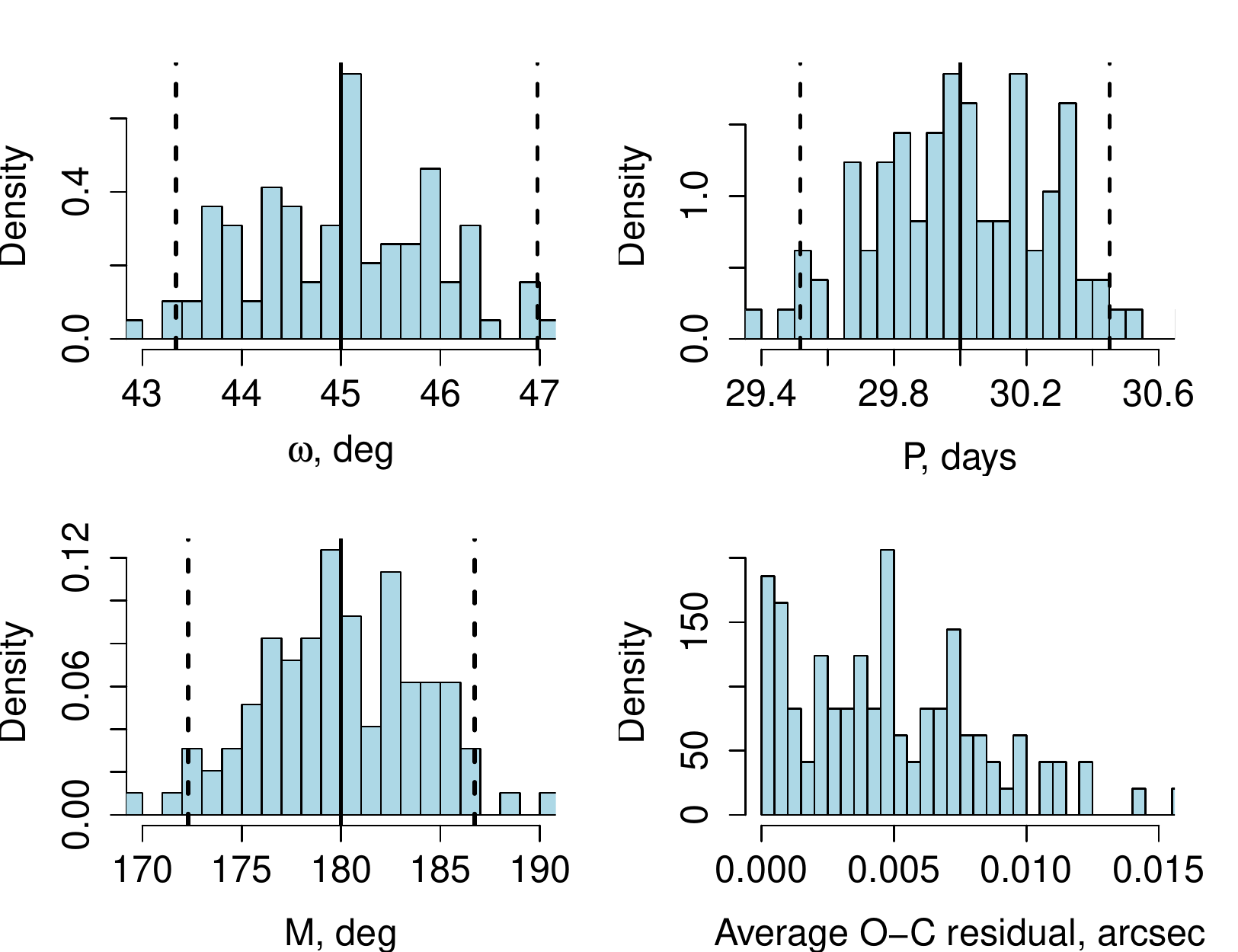}
\caption{Resulting distributions of argument of pericenter $\omega$, mean anomaly $M$ (at 15th simulation day, referenced to J2000 equatorial frame), orbital period $P$ and the average O-C residual obtained for simulated observations with the likelihood model 1. Solid line represents the true parameter value. The doted lines represent the 2.5$\%$ and 97.5$\%$ quantiles of the  resulting sample.}
\label{fig:simMod12}
\end{figure}

For the statistical inference we repeat the SA algorithm 100 times, giving different random initial sets of orbital parameters (see Section \ref{sec:uncertEstimation}). The resulting distribution of orbit samples, obtained with the first model of the likelihood, is shown on Figures \ref{fig:simMod11} -\ref{fig:simMod12}. The distributions obtained with the other three models are similar to those of the first model and are not graphically shown here. According to 
the obtained distributions, we retain the mean value, the standard deviation and the $2.5\%$-$97.5\%$ quantiles for each orbital parameter. This statistical summary for the four likelihood models is listed in Table \ref{tab:resultSim}. The mean values of the obtained distributions (see Table \ref{tab:resultSim}) are very close to the original orbital parameters. This result is obtained despite the fact that the search interval for the parameters, given through the \textit{a priori} term, was chosen as large as possible. This also demonstrates how the MAP converges towards the optimal solution.


The uncertainties from models 1 and 3 are greater than those for models 2 and 4, respectively, following both -- the $2\sigma$ and the inter-quantile -- intervals  estimations. Still, for all the models, there is an agreement between the length of the inter-quantile interval and the $4\sigma_{\theta}$. This indicator together with the symmetric bell-shape of the parameters distributions (e.g. presented by histograms on Figures \ref{fig:simMod11} -\ref{fig:simMod12}) suggest no evidence against un-biasedness and Gaussian statistical behaviour of the obtained results. Since the chosen \textit{a priori} was uniform for all the parameters, this tends to be consistent with the asymptotical properties of the maximum likelihood estimates.

In addition, the average sky plane residual for each orbit sample is done using the following formula: 
\begin{equation}
\overline{\hbox{O-C}} =\dfrac{1}{N}\sum_{i=1}^N\left( |x^{o}_i-x^c_i|^2 +  |y^o_i-y^c_i|^2 \right)^{1/2},
\label{eq:rms}
\end{equation}
The ranges of the average O-C values for the obtained sample of orbits, associated to different likelihood models in the MAP algorithm, are listed in Table \ref{tab:meanOC}. The histogram of the sky plane residuals, shown in Figure~\ref{fig:simMod12}, indicates the prevalence of the rather small values, hence a rather good fitting of the model.

To summarise, no remarkable advantage of any likelihood model over others is noticed. The method converges to the optimal orbit equally with the four models. The outcome of different uncertainties doesn't favor one model over the others.

\begin{table}
\centering
\caption{Range of the average O-C residual, computed with the formula \eqref{eq:rms}, for obtained samples of orbits associated to the four likelihood models.}
\label{tab:meanOC}
\begin{tabular}{ll}
Likelihood & $\min(\overline{\hbox{O-C}})-\max(\overline{\hbox{O-C}})$, arcsec \\ \hline
Model 1    & 0.00006                - 0.0156               \\
Model 2    & 0.0001                 - 0.0128               \\
Model 3    & 0.0005                 - 0.0126               \\
Model 4    & 0.00005                - 0.0126               \\ \hline
\end{tabular}
\end{table}

\subsection{Real observations}
For real data case study the trans-Neptunien binary Teharonhiawako (2001 QT$_{297}$) with its companion named Sawiskera was chosen. The 16 observations published by \cite{2011Grundy} are used. The observations are distributed on the interval from 2001/10/11 to 2010/08/03, covering the entire orbital period (see \cite{2011Grundy}). Some observations were made consecutively with one day interval, but there are also those that follow after one year and more intervals (see Table \ref{tab:obsTeko}). The estimated astrometric errors published with the observations were assumed to follow Gaussian distributions \citep{2011Grundy} with standard deviation magnitudes from 0.13 to 0.003 arcsec.

First, the MAP algorithm for the aforementioned observations is used. Then, the least squares orbit fitting is applied to the same data, using the initial set of parameters from \cite{2011Grundy}.
We will consider the solution obtained with the LS method as a reference in order to verify the result from the MAP method. The reasons we use the LS techniques here are the following. The LS method is less computationally expensive. Although the LS method guarantees only local convergence, in the case of the good initial guess of parameters and the well-distributed data (here both given by \cite{2011Grundy}), the LS method provides an accurate solution, allowing to consider it as a reference. Furthermore, the LS method allows us to easily switch between the not-weighted and the weighted cases. Hence, an appropriate comparison of the first/third and the second/fourth MAP likelihood models with the not-weighted and the weighted LS models, respectively, can be done.

\begin{table}
\centering
\caption{Derived parameters $\hat{\theta}$ with uncertainty estimate for MAP and LS methods obtained from real observations of Teharonhiawako 2001 QT$_{297}$. $\sigma_\theta$ denotes an empirical standard deviation. \textit{m.} denotes a likelihood model of the MAP method or lest-squares (LS) and weighted least-squares (WLS) methods. $\Delta q$ denotes $q(0.975)-q(0.025)$, namely 2.5$\%$ and 97.5$\%$ quantiles interval. 
a) Angles are referenced to J2000 equatorial coordinates. b) The mean anomaly is referred to the epoch 2452000.0 JD. }
\label{tab:results}
\medskip
\begin{tabular}{llll}
Parameter     &  m.   & $\hat{\theta} \pm$ 2 $\sigma_\theta$             & $\Delta q$                     \\ \hline
$a$, km       & 1 & 28100.87 $\pm$ 90.06              & $180.11$                     \\
              & 2 & 28124.49 $\pm$ 1.69              & $3.37$                     \\
              & 3 & 28006.11 $\pm$ 143.91              & $287.82$                     \\
              & 4 & 27777.15 $\pm$ 2.4              & $4.79$                     \\ 
              & LS  & 28125.8 $\pm$ 407.2  &                                             \\
              & WLS & 27780.18 $\pm$ 1.78  &                                             \\ 
              \hline
$e$           & 1 & $0.2439\pm0.0017$   & $0.003$ \\
              & 2 & $0.2435\pm0.00006$   & $0.0001$  \\
              & 3 & $0.249\pm0.00213$   & $0.00373$ \\
              & 4 & $0.2547\pm0.00014$   & $0.00027$  \\               
              & LS  & 0.2436 $\pm$ 0.0081  &                                             \\
              & WLS & 0.2548 $\pm$ 4e-5    &                                             \\
               \hline
$i^a$, deg      & 1 & $145.49\pm0.26$                       & $0.45$                     \\
              & 2 & $144.01\pm0.01$                       & $0.02$                     \\
              & 3 & $145.37\pm0.44$                       & $0.74$                     \\
              & 4 & $143.98\pm0$                       & $0.01$                     \\ 
              & LS  & 144.01 $\pm$ 1.5     &                                             \\
              & WLS & 143.99 $\pm$ 0.01    &                                             \\
 \hline
$\Omega^a$, deg & 1 & $54.28\pm0.71$                     & $1.3$                   \\
              & 2 & $51.88\pm0.02$ & $0.04$                   \\
              & 3 & $55.11\pm0.42$                     & $0.77$                   \\
              & 4 & $55.08\pm0.02$ & $0.07$                   \\ 
              & LS  & 51.89 $\pm$ 2.5      &                                             \\
              & WLS & 55.09 $\pm$ 0.01     &                                             \\
 \hline
$\omega^a$, deg & 1 & $324.97\pm0.71$                      & $1.3$                    \\
              & 2 & $324.32\pm0.03$  & $0.05$                    \\
              & 3 & $325.65\pm0.2$                      & $0.39$                    \\
              & 4 & $324.81\pm0.02$                      & $0.04$                    \\                        & LS  & 324.34 $\pm$ 2.5     &                                             \\
              & WLS & 324.84 $\pm$ 0.01    &                                             \\
 \hline
$P$, days    & 1 & $828.21\pm0.36$                       & $0.7$                     \\
              & 2 & $828.17\pm0.01$                       & $0.03$                     \\
              & 3 & $828.4\pm0.24$                       & $0.39$                     \\
              & 4 & $828.07\pm0.02$  & $0.04$                     \\ 
              & LS  & 828.15 $\pm$ 80.9    &                                             \\
              & WLS & 828.07 $\pm$ 0.004   &                                             \\           
              \hline
M$^b$, deg    & 1 & $276.02\pm0.24$                       & $0.46$                     \\
              & 2 & $275.27\pm0.02$                       & $0.03$                     \\
              & 3 & $276\pm0.18$                       & $0.38$                     \\
              & 4 & $276.7\pm0.02$                       & $0.03$                     \\ 
              & LS  & 275.26 $\pm$ 1.5     &                                             \\
              & WLS & 276.68 $\pm$ 0.01    &                                             \\               		
              \hline
\end{tabular}
\end{table}

\paragraph*{Results.} 

In the same way as for the simulated observations, for the real data, we obtain the distributions of each orbital parameter through 100 SA evaluations. For these distributions, we summarise the inter-quantile interval $q(0.025)-q(0.975)$ in Table \ref{tab:results}. The set of parameters $\hat{\btheta}$ of the best-fitted orbit with $2\sigma_{\theta}$ uncertainties and the LS result are also summarised in Table \ref{tab:results}. 

The results obtained for the real data using the SA algorithm are similar to the ones obtained for the simulated data. Again, the distribution shape, the numerical values of the $\sigma_\theta$ and the length of the inter-quartile interval, do not indicate an obvious bias, and the parameters marginals distributions tend to exhibit a Gaussian behaviour.

It can be noticed that the result, obtained with the first likelihood model, is slightly shifted from the LS result. The likelihood model 2 converges practically to the same result as the not-weighted LS. The slightly different results of the model 1 and the model 2 (or LS) are not contradictory, but are explained by the different likelihood models. Concerning the weighted LS, the results have the same tendency than the not-weighted models: model 3 obtains resulting distributions shifted from the LS results and model 4 provides results very close to those of the LS (see Table \ref{tab:results}).

Under assumption, that the LS method provides the true solution, the fact that the MAP algorithm gives results similar to the LS method validates the new method. Hereafter, we retain the second (for not-weighted observations) and the fourth (for weighted observations) likelihood models being more appropriate, since these models provide the result closer to those of the LS method (in the not-weighted and the weighted cases, respectively).

It should be noticed that in case of well-distributed observations and good initial estimation of parameters, the LS method appears to be more practical, since it is less computationally expensive than the new proposed method. Hence, this particular situation doesn't require the use of our method; the classical LS based method can be applied in order to save computation time. However, there is not a rigorous criteria to state whether the data and initial guesses of parameters are good enough. Under this circumstance, the MAP method is more universal.

\subsection{Ephemeris prediction}

\begin{figure}
\centering
\includegraphics[width=\linewidth]{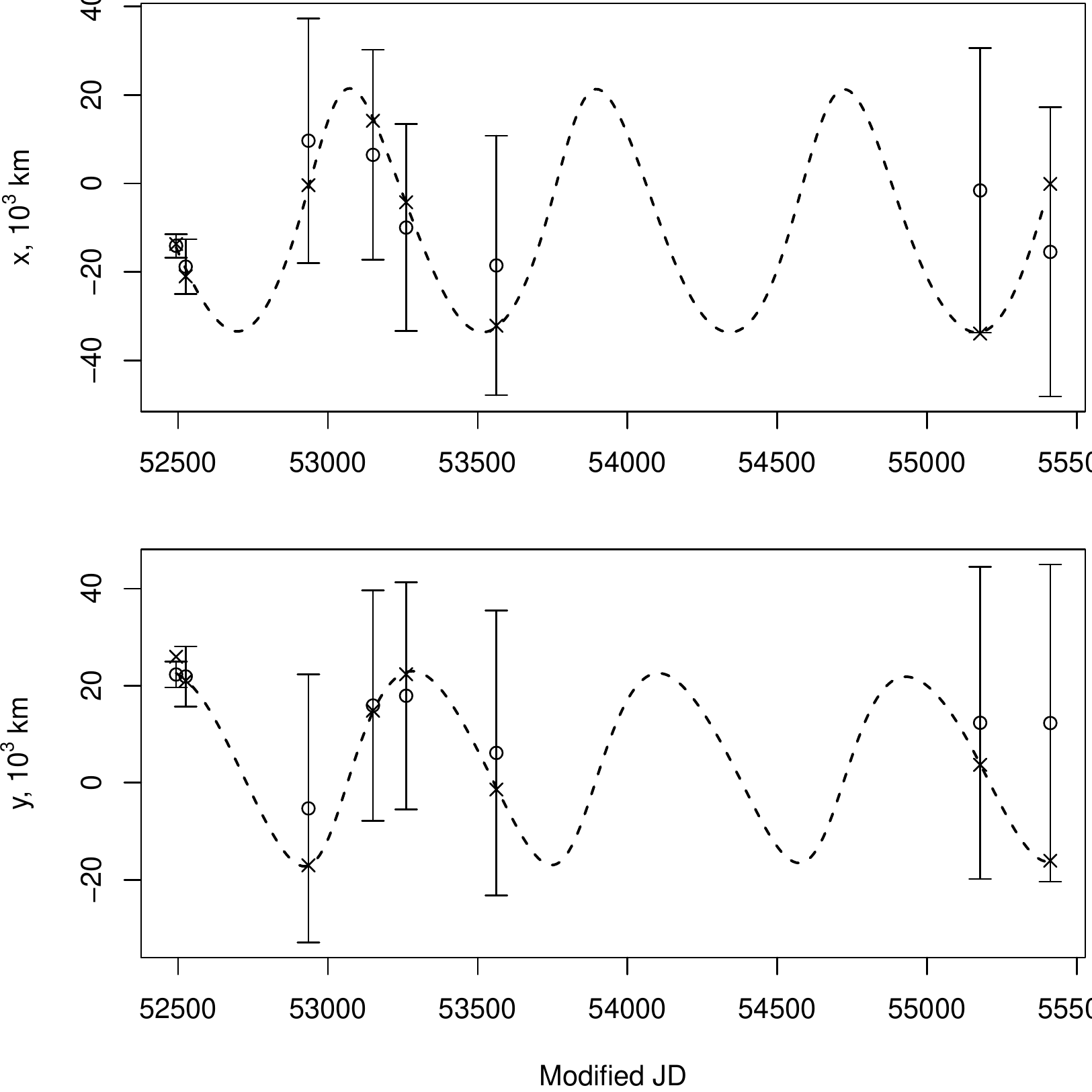}
\caption{Ephemeris prediction for the result of MAP method with the likelihood model $2$. The calculated positions (black circles) are compared with given observed positions (crosses) by the $x$ and $y$ coordinates on sky-plane. Black bars denote the 2.5$\%$-97.5$\%$ quantiles interval. Dotted line corresponds to the calculated positions for the orbit, obtained with the entire set of observations.}
\label{fig:unc2CI}
\end{figure}

\begin{figure}
\centering
\includegraphics[width=\linewidth]{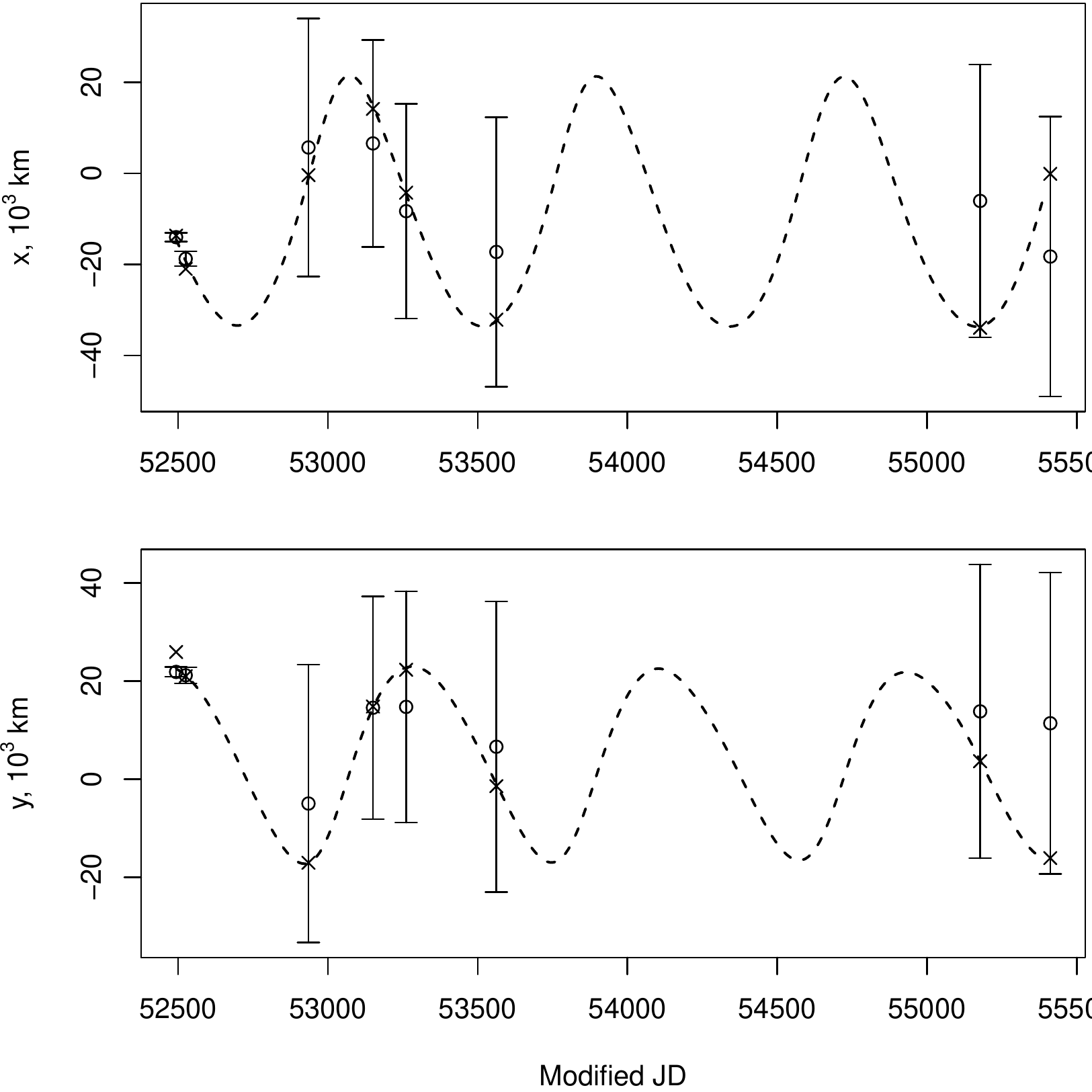}
\caption{Ephemeris prediction for the result of MAP method with the likelihood model $4$. The calculated positions (black circles) are compared with given observed positions (crosses) by the $x$ and $y$ coordinates on sky-plane. Black bars denote the 2.5$\%$-97.5$\%$ quantiles interval. Dotted line corresponds to the calculated positions for the orbit, obtained with the entire set of observations.}
\label{fig:unc4CI}
\end{figure}

In order to verify the capability of the MAP method to make ephemeris prediction, we apply it to a reduced set of observations. Namely, the eight first observations from the 16 given in \cite{2011Grundy} are used for the orbit determination problem. It should be noted that for this reduced set of observations, the LS method does not converge, whereas the MAP method provides a solution. Using this solution we calculate the ephemeris at the epochs of the other eight observations following after the first set. Following the method presented in Section \ref{sec:uncertEstimation}, 95$\%$ inter-quantile intervals were computed for the calculated positions.

Figures \ref{fig:unc2CI} and \ref{fig:unc4CI}, associated to the second and fourth likelihood models, respectively, show observed and predicted positions together with the 95$\%$ inter-quantile interval. It can be seen, for both likelihood models, that the estimated uncertainties can be as large as the apparent size of orbit (see the $x$ and $y$ amplitudes on the figures). The test of the validity of the model is rejected only by one observation: the 9th (see the $y$ coordinate of the first observed position on Figures \ref{fig:unc2CI} and \ref{fig:unc4CI}), that can be caused by an observational bias or the imperfection of the chosen physical model. Still, all of the other observed positions are situated within the range of the estimated interval. Hence, we may consider the observed positions as realisations of the stochastic process providing the different MAP solutions.

\section{Conclusion and perspectives}
\label{sec:conclusions}
The orbital parameters estimation by the MAP with the simulated annealing algorithm has been proposed for the binary asteroid orbit determination problem. The new method has been successfully applied to simulated and real observations and verified for ephemeris prediction. The developed algorithm is prepared to be implemented into the Gaia mission \citep{prusti2016gaia} data reduction pipeline.

The MAP method is independent of the initial orbit estimation, guaranteeing the theoretical convergence towards the global optimum solution. This advantage allows us to use it for newly discovered asteroids, when an initial parameters estimation is difficult. At the same time, if any available information helps to constrain the parameters search space, it can be easily introduced through the \textit{a priori} term.

The new method enables us to consider and to compare different forms of observational errors distribution implementing it by the likelihood modelling. This aspect can be particularly useful when the observational errors cannot be assumed to obey Gaussian distribution.

The sampling  mechanism within the simulated annealing algorithm is done through the orbital parameters. This approach simplifies the entire procedure, since no methods (such as the Thiele-Innes, or equivalent) are required for computation of orbital parameters from observations. 

The parameters estimation by the MAP with the SA algorithm implementation is still developing. The method has been implemented for only a few binaries (see also \cite{2016IAUSKovalenko}) and we expect to improve it when applied in more various cases. For this purpose, the following perspectives are depicted. First, even if guaranteeing the convergence of the method and the quality of the obtained results, the choices of the cooling schedule and of the proposal distribution in the SA algorithm are not optimal. Second, the dynamical model of a binary motion can be completed considering perturbed orbit instead of keplerian. Third, the new method can be expanded for the non-resolved binary asteroids.

\section*{Acknowledgements}
This work is supported by Labex ESEP (ANR N 2011-LABX-030) and by the Russian Foundation for Basic Research, project no. 16-52-150005-CNRS-a. The work of the second author was done while he was working at Universit\'e de Lille - Laboratoire Paul Painlev\'e.




\bibliographystyle{mnras}
\bibliography{biblio} 




\appendix

\section{Simulated and real observations}
\label{app:A}
\begin{table}
\centering
\caption{Simulated observations: $x = (\alpha_2-\alpha_1)\cos\delta_1$ and $y=\delta_2-\delta_1$,  where $\alpha$ is right ascension, $\delta$ is declination, refereed to primary (1) and secondary (2), respectively; $\alpha$ is right ascension and $\delta$ is declination of the target asteroid; $R$ is the distance from the observer to the target. Estimated $\sigma_{x,i}$ and $\sigma_{y,i}$ uncertainties in the final digits are indicated in parentheses. }
\medskip
\label{tab:obssim}
\begin{tabular}{ccclll}
Day & \multicolumn{2}{c}{$x$ and $y$ (arcsec)} & R (AU) & $\alpha$ (deg) & $\delta$ (deg)\\ \hline
5 & -0.0072(10) & 0.1741(10) & 44.87 & 56.02 & 24.01 \\
14 & 0.2663(20) & -0.0662(10) & 44.74 & 56.18 & 24.01 \\
26 & 0.1413(30) & -0.2143(40) & 44.60 & 56.02 & 24.01 \\
30 & -0.0939(40) & 0.0318(10) & 44.56 & 56.21 & 24.01 \\
38 & 0.1068(50) & 0.1118(40) & 44.49 & 56.07 & 24.01 \\
41 & 0.2004(90) & 0.0248(10) & 44.46 & 56.02 & 24.00 \\
47 & 0.3012(10) & -0.1494(40) & 44.43 & 55.15 & 24.02 \\
50 & 0.2989(70) & -0.2142(10) & 44.41 & 55.09 & 24.01 \\
59 & -0.0405(10) & -0.0545(40) & 44.39 & 55.16 & 24.01 \\
62 & -0.1082(80) & 0.1558(10) & 44.38 & 55.10 & 24.01\\ \hline
\end{tabular}
\end{table}

\begin{table}
\centering
\caption{Observations of (2001 QT$_{297}$) Teharonhiawako \citep{2011Grundy}, where the relative $x = (\alpha_2-\alpha_1)\cos\delta_1$ and  $y=\delta_2-\delta_1$,  with $\alpha$ and $\delta$ are right ascension and declination respectively, and subscripts 1 and 2 refer to primary and secondary, respectively. The distance from the observer to the target is $R$. Estimated Estimated $\sigma_{x,i}$ and $\sigma_{y,i}$ uncertainties in the final digits are indicated in parentheses.}
\medskip
\label{tab:obsTeko}
\begin{tabular}{lllll}
UT day     & and hour & $R$ (AU) & x(arcsec)    & y(arcsec)    \\ \hline
11/10/2001 & 0.9528    & 44.370   & +0.5390 (51) & -0.2770 (52) \\
12/10/2001 & 1.873     & 44.385   & +0.5460 (81) & -0.2675 (84) \\
01/11/2001 & 0.7697    & 44.699   & +0.624  (21) & -0.214  (24) \\
02/11/2001 & 0.4299    & 44.716   & +0.644  (21) & -0.184  (26) \\
03/11/2001 & 0.3249    & 44.733   & +0.642  (40) & -0.193  (39) \\
04/11/2001 & 0.9046    & 44.750   & +0.645  (21) & -0.138  (35) \\
13/07/2002 & 6.7387    & 44.130   & -0.314  (23) & +0.692  (29) \\
18/07/2002 & 6.9538    & 44.085   & -0.344  (68) & +0.700  (55) \\
07/08/2002 & 4.5629    & 43.970   & -0.43   (13) & +0.81   (13) \\
08/09/2002 & 5.7632    & 44.024   & -0.658  (91) & +0.658  (91) \\
23/10/2003 & 1.7567    & 44.536   & -0.012  (60) & -0.527  (50) \\
25/05/2004 & 8.789     & 44.878   & +0.4350 (70) & +0.4560 (70) \\
13/09/2004 & 3.2531    & 44.074   & -0.1330 (70) & +0.6990 (60) \\
11/07/2005 & 5.8782    & 44.234   & -1.0020 (80) & -0.0440 (80) \\
12/12/2009 & 5.207     & 45.368   & -1.0257 (77) & +0.1098 (53) \\
03/08/2010 & 10.1942   & 44.166   & -0.0032 (30) & -0.5015 (30) \\ \hline\end{tabular}
\end{table}


\bsp	
\label{lastpage}
\end{document}